# Automatic 8-tissue Segmentation for 6-month Infant Brains

Yilan Dong[1,2] [0000-0001-9731-8362], Vanessa Kyriakopoulou[1,2][0000-0002-9883-0314], Irina Grigorescu[1][0000-0002-9756-3787], Grainne McAlonan[2][0000-0002-4466-2343], Dafnis Batalle[1,2][0000-0003-2097-979X], Maria Deprez[1] [0000-0002-2799-6077]

[1] School of Biomedical Engineering & Imaging Sciences, King's College London, London SE1 7EH, United Kingdom
[2] Department of Forensic and Neurodevelopmental Science, Institute of Psychiatry, Psychology & Neuroscience, King's College London, London SE5 8AF, United Kingdom
✉ yilan.dong@kcl.ac.uk

**Abstract.** Numerous studies have highlighted that atypical brain development, particularly during infancy and toddlerhood, is linked to an increased likelihood of being diagnosed with a neurodevelopmental condition, such as autism. Accurate brain tissue segmentations for morphological analysis are essential in numerous infant studies. However, due to ongoing white matter (WM) myelination changing tissue contrast in T1- and T2-weighted images, automatic tissue segmentation in 6-month infants is particularly difficult. On the other hand, manual labeling by experts is time-consuming and labor-intensive. In this study, we propose the first 8-tissue segmentation pipeline for six-month-old infant brains. This pipeline utilizes domain adaptation (DA) techniques to leverage our longitudinal data, including neonatal images segmented with the neonatal Developing Human Connectome Project structural pipeline. Our pipeline takes raw 6-month images as inputs and generates the 8-tissue segmentation as outputs, forming an end-to-end segmentation pipeline. The segmented tissues include WM, gray matter (GM), cerebrospinal fluid (CSF), ventricles, cerebellum, basal ganglia, brainstem, and hippocampus/amygdala. Cycle-Consistent Generative Adversarial Network (CycleGAN) and Attention U-Net were employed to achieve the image contrast transformation between neonatal and 6-month images and perform tissue segmentation on the synthesized 6-month images (neonatal images with 6-month intensity contrast), respectively. Moreover, we incorporated the segmentation outputs from Infant Brain Extraction and Analysis Toolbox (iBEAT) and another Attention U-Net to further enhance the performance and construct the end-to-end segmentation pipeline. Our evaluation with real 6-month images achieved a DICE score of 0.92, an HD95 of 1.6, and an ASSD of 0.42.

# 1   Background

In recent years, there has been increasing interest in infant brain development, characterized by rapid brain growth, and evolving cognitive and motor functions. Neurobiological findings in early childhood suggest that early brain overgrowth may be associated with autism [1]. A recent study has also shown autism phenotypes are associated with variations in white matter development [2]. These studies illustrate the significance of acquiring volumetric measurements in infants, emphasizing the importance of such assessments in understanding developmental trajectories and potential correlations with various neurological and neurodevelopmental conditions.

Magnetic resonance imaging (MRI), as an advanced non-invasive and high-resolution imaging technique, has underpinned remarkable advancements in the medical field. However, it faces challenges during the 6-month period of infancy. One primary challenge is the low tissue contrast between GM and WM due to rapid myelination development [3], making them difficult to distinguish on the MRI scans.

Accurate brain tissue segmentations for morphological analysis are essential in infant studies, but manual labeling by experts is time-consuming and labor-intensive. Most existing image preprocessing software, such as FSL, is primarily designed for adult brain segmentation and performs poorly on infant brain images due to low tissue contrast and poor image quality [4]. Although the Infant Brain Extraction and Analysis Toolbox (iBEAT) [5] is widely used in infant research for its efficiency in processing and analyzing infant brain, it solely performs WM, GM and CSF segmentations.

Machine learning (ML) based solutions have also been proposed, with segmentation algorithms achieving state-of-the-art performance on six-month infant brain datasets in the iSeg-2017 and iSeg-2019 challenges [3, 6]. However, no participants have achieved consistent performance across six-month infant datasets from multiple sites. Recently, domain adaptation (DA) techniques have been suggested as a potential method to improve performance in this task [6].

DA methods have gained increasing interest in the medical imaging field to mitigate the distribution gap between training and test datasets. Generative adversarial network (GAN) and its extensions, such as CycleGAN [7], are widely employed for image-to-image translation to achieve image domain transformation between the target and source domains. A study proposed a 3D CycleGAN-Seg network that leveraged 24-month annotation to segment 6-month images [8]. Another study utilized annotation from the 12-month images to segment 6-month images using the proposed semantics-preserved GAN, and Transformer based multi-scale segmentation network [9]. However, all these studies focused solely on WM, GM and CSF segmentation. Other brain tissues, including the ventricles, hippocampus, and cerebellum, also play crucial roles in infant brain research.

In this paper, we introduce the first DA-based 8-tissue brain segmentation pipeline for 6-month infants utilizing the annotation information from neonatal images processed with the developing Human Connectome (dHCP neonatal structural pipeline [10]). Tissue segmentations available as part of the dHCP pipeline include: CSF, GM, WM, ventricle, cerebellum, basal ganglia, brainstem, and hippocampus/amygdala. To obtain the best segmentation performance, we designed different combinations of CycleGAN, Attention UNet, Voxelmorph, and iBEAT outputs and compared their performances on real 6-month images. The highest segmentation performance was achieved by the combination of CycleGAN, Attention UNet, and iBEAT, reaching a DICE score of 0.92, an HD95 of 1.6, and an ASSD of 0.42 on the real 6-month T1w and T2w images.

## 2    Methodology

### 2.1    Data Acquisition and Preprocessing

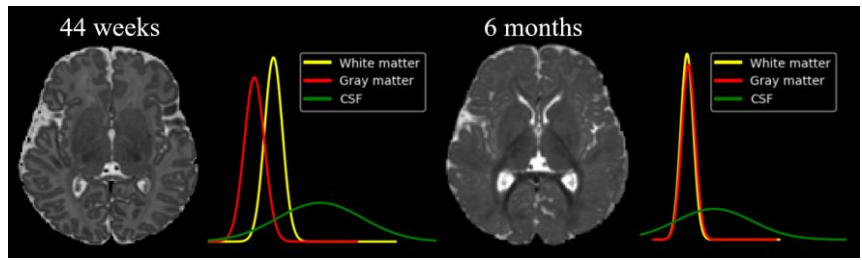

**Fig. 1** The neonatal and 6-month images and corresponding histogram distributions of WM, GM and CSF.

**Participants:** 43 subjects participated in this study as part of the Brain Imaging in Babies (BIBS) project. The infants were born between 34 and 42 weeks of gestation, comprising 22 males and 21 females in the cohort. All participants were scanned as both neonates (see "Neonatal scans" below) and as 6-month-olds (see "6-month-old infant scans" ).

**Neonatal scans:** T2w neonatal images were scanned between 37 and 44 weeks postmenstrual age (PMA). Images were acquired on a Philips Achieva 3T scanner equipped with a dedicated neonatal brain imaging system (NBIS) and a 32-channel neonatal head coil. T2w images were obtained using a T2w turbo spin echo (TSE) with fat suppression, TR = 12000 $ms$, TE = 156 $ms$, a resolution of $0.8 \times 0.8$ $mm^2$ and a slice thickness of 1.6 $mm$.

**6-month-old infant scans:** T1w and T2w 6-month images were scanned between 5 months to 7.7 months. Images were acquired on a 3T Phillips Scanner employed with a 32-channel adult head coil. T2w images were obtained using Spin Echo with TR = 15000 $ms$, TE = 120 $ms$, a resolution of $0.86 \times 0.86$ $mm^2$ and a slice thickness of 2 $mm$. The T1w images were acquired using Gradient Echo (GE) with TR = 12 $ms$, echo time TE = 4.6 $ms$, a resolution of $0.78 \times 0.78$ $mm^2$ and a slice thickness of 1.6 $mm$. **Fig. 1** visualizes a comparison between 44 weeks neonatal and 6-month infant images, along with their corresponding WM, GM and CSF histogram distributions.

**Data preprocessing:** We split the 43 participants into a training set ($n = 33$) and a test set ($n = 10$). The dHCP neonatal image preprocessing pipeline was applied to all 43 neonatal images, including motion correction, super-resolution reconstruction and tissue segmentation [11]. We utilized the 8 brain tissue segmentation outputs for subsequent experiments (excluding the "skull" output). The remaining segmentation outputs CSF, GM, WM, ventricle, cerebellum, basal ganglia, brainstem, hippocampus/amygdala were used in the subsequent experiments.

For 6-month infant data, we designed a custom preprocessing pipeline that included N4 bias correction, registration, separate brain skull removal for T1w and T2w images using machine learning algorithm, cropping and resampling.

## 2.2   Machine learning models in pipelines

To investigate the best solution for segmenting our target dataset (6-month data), we train and compare five different DL pipelines:

**AUNet (baseline):** A MONAI Attention UNet [12] was trained on the neonatal T2w images and labels, then applied directly to real 6-month T2w images.

**Cyc+AUNet:** CycleGAN was employed to transform neonatal T2w images into synthesized 6-month T2w images (neonatal images with 6-month intensity contrast). At the same time, an Attention UNet was trained on these synthesized 6-month images to predict their corresponding neonatal labels.

**Cyc+AUNet+VM:** Using the pre-trained **Cyc+AUNet**, we further employed VoxelMorph [13] to register synthesized 6-month images to the real 6-month image space (paired), and keep training the Attention UNet on the warped synthesized 6-month images to predict warped neonatal labels.

**Cyc+AUNet+iBEAT:** We employed the same strategy as for Cyc+AUNet, but replaced the WM, GM and CSF segmentation outputs of **Cyc+AUNet** with iBEAT segmentation outputs.

**Cyc+AUNet+iBEAT+AUNet:** Finally, we trained a second Attention UNet on the real 6-month images and their segmentation outputs from **Cyc+AUNet+iBEAT**.

**Fig. 2** illustrates how CycleGAN, Attention UNet, VoxelMorph and iBEAT cooperate with each other.

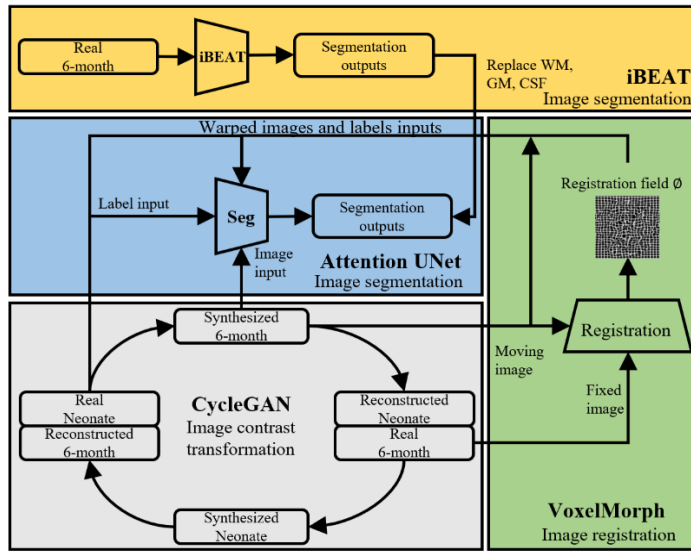

**Fig. 2**. The cooperation between CycleGAN, Attention UNet, VoxelMorph and iBEAT

**Network architectures:**

**Attention UNet (AUNet)** is an extension of the traditional U-Net architecture, which incorporates attention mechanisms to enhance the model's ability to focus on relevant features. The model parameters were optimized using a combination of DICE loss and Cross Entropy loss (CE), by minimizing the following loss function:

$$Loss_{seg} = Loss_{DICE}(AUNet(X), Y) + Loss_{CE}(AUNet(X), Y) \qquad (1)$$

**CycleGAN (Cyc).** The primary goal of CycleGAN is to learn mappings between two different image domains in an unsupervised manner, which is particularly useful in scenarios where obtaining paired data is difficult or expensive. The architecture of CycleGAN involves two generators, $G_{AB}$ and $G_{BA}$, along with two discriminators, $D_A$ and $D_B$. Here, we define the domain $A$ as the neonatal domain and the domain $B$ as the 6-month domain. Generator $G_{AB}$ learns the mapping from neonatal images to 6-month images, such as the transformation from real neonatal images $X_{neo}^{real}$ to synthesized 6-month images $X_{6-month}^{syn}$. Conversely, generator $G_{BA}$ learns the inverse mapping from 6-month images to neonatal images. The discriminators $D_A$ and $D_B$ are responsible for distinguishing between real images and synthesized images. The objective function of CycleGAN in **Fig. 2** is defined as:

$$Loss_{Cyc} = Loss_{adversarial}(G_{AB}, D_B; G_{BA}, D_A) + Loss_{consistency}(G_{AB}, G_{BA})$$
$$+ Loss_{identity}(G_{AB}, G_{BA}) + Loss_{seg}\left(Seg(X_{6-month}^{syn}), Y_{neo}\right) \qquad (2)$$

where the $Loss_{adversarial}$ encourages the generated images to become indistinguishable from the target domain, thereby facilitating domain translation and image synthesis. The $Loss_{consistency}$ and $Loss_{identity}$ encourage the bi-directional transformation consistency and the preservation of important features during the translation process. The loss function from Attention UNet was also included to contribute to the generation of synthesized 6-month images $X_{6-month}^{syn}$.

**VoxelMorph (VM)** [13] registration model was employed to register the synthesized 6-month images to the real 6-month images. This is because the contrast transferred data (synthesized 6-month images) retained the neonatal cortical folding pattern. This, in turn, has the potential to reduce the segmentation performance on the real 6-month images, as the segmentation model is trained solely on images with simpler neonatal cortical folding compared to real 6-month images. In this work, the input to VoxelMorph are pairs of 3D moving ($M$) and fixed ($F$) images, where $M$ corresponding to the neonatal data, while $F$ corresponds to the 6-month data. Specifically, VoxelMorph is trained to produce a deformation field $\emptyset$, which warps $M$ and its corresponding neonatal labels $Y_{neo}$ to obtain the warped moving image $M(\emptyset)$ and labels $Y_{neo}(\emptyset)$ as the new inputs for the Attention UNet. The objective function is defined as the combination of the local normalized cross correlation loss (LNCC, $Loss_{LNCC}$) and a regularization term $Loss_{smooth}$.

$$Loss_{Vox} = Loss_{LNCC} + \lambda * Loss_{smooth}$$
$$= \frac{1}{N_s} \frac{\sum_{i \in N_s}(F(i) - \bar{F}) \cdot \left(M(\emptyset(i)) - \overline{M(\emptyset)}\right)}{\sqrt{\sum_{i \in N_s}(F(i) - \bar{F})^2 \cdot \sum_{i \in N_s}\left(M(\emptyset(i)) - \overline{M(\emptyset)}\right)^2}}$$
$$+ \sum_{i \in \Omega} \|\nabla \emptyset(i)\|^2 \qquad (3)$$

where $N$ is the total number of voxels in the fixed image $F$, $i$ denotes the location of each voxel, and $s$, the sliding window size of LNCC, is set to 15. $F(i)$ and $M(\emptyset(i))$ represent the intensity values at location $i$ in $F$ and $M(\emptyset)$. $\bar{F}$ and $\overline{M(\emptyset)}$ denote the mean intensity values of the fixed and transformed moving images. The weight $\lambda$ for $Loss_{smooth}$ is set to 0.6 based on the best segmentation performance obtained from a grid search of different $\lambda$ values (0.1, 0.2,…0.9).

**iBEAT** [5] segmentation pipeline serves as an alternative method to address the challenge posed by the complex cortical patterns in 6-month images. We fed real 6-month T1w and T2w images, along with brain masks, into iBEAT to obtain segmentation outputs for WM, GM and CSF. These outputs from iBEAT were then utilized to replace the WM, GM, and CSF segmentation outputs from **Cyc+AUNet**.

### 2.3    Model implementation details

The 3D Attention UNet from Project Monai [14] was implemented with 5 encoder-decoder blocks featuring 32, 64, 128, 256, and 512 filters, using a kernel size of 3 and a stride of 2, with a learning rate of 0.0004.

For the 3D CycleGAN generators, UNet with 7 layers was employed, with channel configurations of 64, 128, 256, 512, and 512. The discriminator comprised a 5 layers PatchGAN, with filters of 64, 128, 256, 512, and 1. Convolutional layers in CycleGAN had a kernel size of 3, a stride of 2, and were optimized with a learning rate of 0.0008.

The 3D VoxelMorph model consisted of 4 convolutional layers with filter sizes of 16, 32, 32, and 32 [15]. These layers utilized a kernel size of 3, a stride of 2, and were optimized with a learning rate of 0.002.

All model parameters were optimized using the Adam optimizer, and experiments were conducted on the NVIDIA A100 Tensor Core GPU.

## 3    Results

### 3.1    Test set and evaluation criteria

We utilized real 6-month images and manually corrected segmentation from 10 individuals as the test set to evaluate segmentation performance of different pipelines. Using the same evaluation criteria, we calculated MONAI's implementation [14] of the DICE score, the 95th percentile Hausdorff distance (HD95) and the average symmetric surface distance (ASSD) for each brain tissue. The results are presented in **Table 1**. The segmented outputs of different pipelines are visualized in **Fig. 3** and **Fig. 4**.

### 3.2    Quantitative comparison

The **AUNet**, as the baseline, was trained on the neonatal images and labels and performed poorly on the real 6-month images, particularly in WM and GM segmentations. It obtained an average DICE score of 0.74, an HD95 of 15.49 and an ASSD of 3.11 for whole brain segmentation (see **Table 1**). After applying image contrast transfer, the **Cyc+AUNet** pipeline showed significant improvement (two-tailed t-test, p=0.046) across most brain tissues compared to the baseline. The average DICE score increased to 0.84, while HD95 and ASSD decreased to 3.07 and 0.66, respectively (see **Table 1**).

**Table 1.** DICE, HD95, ASSD scores of 8 brain tissues in different pipelines.

| Brain Tissue | 1. AUNet | | | 2. Cyc+AUNet | | | 3. Cyc+AUNet+VM | | |
|---|---|---|---|---|---|---|---|---|---|
| | DICE | HD95 | ASSD | DICE | HD95 | ASSD | DICE | HD95 | ASSD |
| CSF | 0.78 | 2.83 | 0.59 | 0.79 | 1.57 | 0.48 | 0.82 | 1.56 | 0.43 |
| GM | 0.73 | 1.94 | 0.68 | 0.73 | 1.63 | 0.58 | 0.77 | 1.45 | 0.55 |
| WM | 0.53 | 3.00 | 1.20 | 0.68 | 2.24 | 0.79 | 0.71 | 2.38 | 0.76 |
| Ventricle | 0.71 | 3.95 | 1.06 | 0.84 | 8.05 | 0.90 | 0.82 | **2.60** | 0.63 |
| Cerebellum | 0.93 | 11.62 | 1.46 | 0.94 | 3.23 | 0.81 | 0.94 | **2.02** | 0.71 |
| Basal Ganglia | 0.70 | 27.06 | 5.82 | 0.94 | 1.9 | 0.56 | 0.94 | 1.60 | 0.57 |
| Brainstem | 0.79 | 47.46 | 10.86 | 0.94 | 4.18 | 0.60 | 0.93 | **1.29** | 0.45 |
| Hippocampus /Amygdala | 0.73 | 26.12 | 3.24 | 0.84 | 1.79 | 0.59 | 0.85 | **1.49** | **0.55** |
| Average | 0.74 | 15.49 | 3.11 | 0.84 | 3.07 | 0.66 | 0.85 | 1.80 | 0.58 |

| Brain Tissue | 4. Cyc+AUNet+iBEAT | | | 5. Cyc+AUNet+iBEAT+AUNet | | |
|---|---|---|---|---|---|---|
| | DICE | HD95 | ASSD | DICE | HD95 | ASSD |
| CSF | 0.80 | 1.56 | 0.49 | **0.93** | **1** | **0.19** |
| GM | 0.86 | 1 | 0.42 | **0.92** | **1** | **0.27** |
| WM | 0.87 | 1.08 | 0.41 | **0.91** | **1** | **0.30** |
| Ventricle | 0.84 | 8.05 | 0.90 | **0.87** | 2.97 | **0.52** |
| Cerebellum | 0.94 | 3.23 | 0.81 | **0.95** | 2.31 | **0.68** |
| Basal Ganglia | 0.94 | 1.90 | 0.56 | **0.95** | **1.25** | **0.43** |
| Brainstem | 0.94 | 4.18 | 0.60 | **0.95** | 1.38 | **0.39** |
| Hippocampus /Amygdala | 0.84 | 1.79 | 0.59 | **0.84** | 1.91 | 0.59 |
| Average | 0.88 | 2.85 | 0.60 | **0.92** | **1.60** | **0.42** |

To further improve segmentation accuracy for WM and GM, we first added a Voxelmorph registration step. The **Cyc+AUNet+VM** pipeline shows that registration improved the DICE scores of WM and GM from 0.73 to 0.77 and from 0.68 to 0.71 respectively. Another method, **Cyc+AUNet+iBEAT**, replaced the WM, GM and CSF outputs with those from iBEAT, resulting in DICE scores for WM and GM increasing to around 0.87, with HD95 and ASSD decreasing to around 1 and 0.4, respectively.

Although adding registration algorithm to the pipeline underperforms compared to **Cyc+AUNet+iBEAT**, it slightly increased the DICE scores of WM, GM and CSF, and significantly reduced the HD95 scores of ventricles, cerebellum and brainstem. The **Cyc+AUNet+iBEAT+AUNet**, performing segmentation on real 6-month T1w and T2w images, exhibited the highest DICE score of 0.92, the lowest HD95 of 1.6, and the lowest ASSD of 0.42, demonstrating the best overall segmentation performance.

We also conducted an ablation study that we trained the **Cyc+AUNet+iBEAT+AUNet** pipeline using 6-month T1w and T2w images, separately. the performance decreased compared to using both modalities at the same time, obtaining a DICE of 0.89, an HD95 of 1.54 and an ASSD of 0.48 on T1 modality, and a DICE of 0.90, an HD95 of 1.55 and an ASSD of 0.46 on T2 modality.

### 3.3  Qualitative assessment

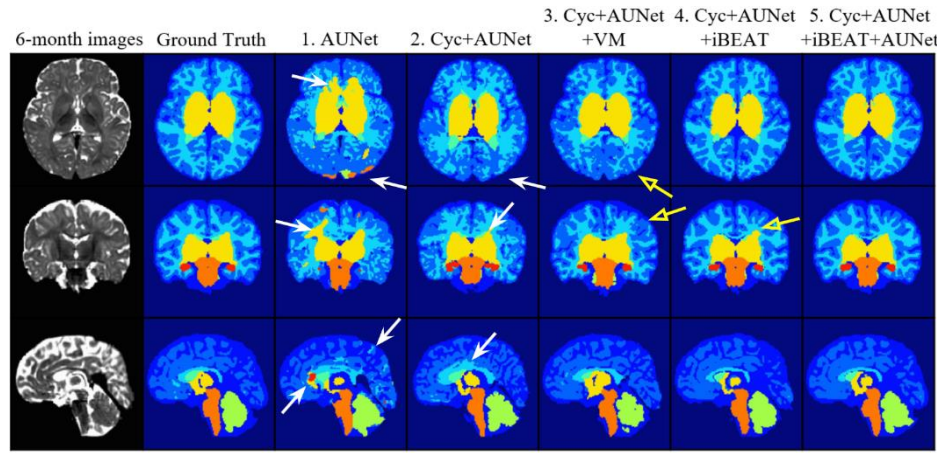

**Fig. 3** The tissue segmentation between different pipelines. The white arrows indicate the gross misclassification in pipeline 1 and 2. Pipeline 3 and 4 performed better, though some misclassifications remain, as indicated by the yellow arrows.

**Fig. 3** visualizes the 8-tissue segmentation comparison across different pipelines. Due to the image contrast transformation provided by CycleGAN, the segmentation of the basal ganglia, brainstem and hippocampus/amygdala is more accurate compared to **AUNet**. Adding registration to **Cyc+AUNet** resulted in less noisy WM segmentation, though it underperforms compared to **Cyc+AUNet+iBEAT**.

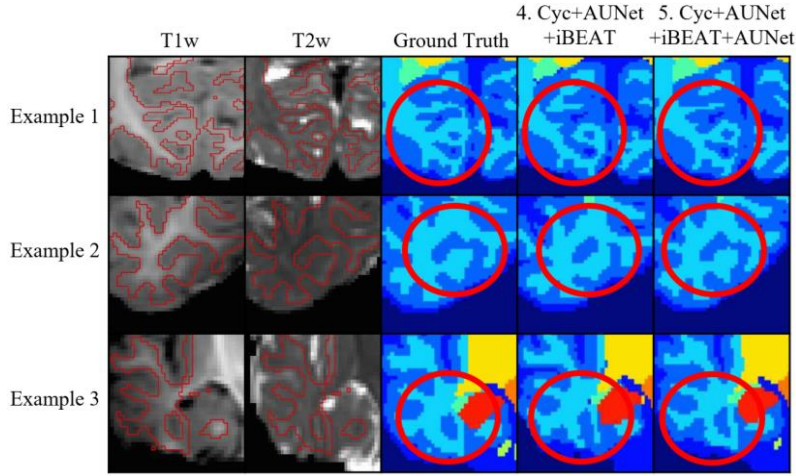

**Fig. 4** The differences between **Cyc+AUNet+iBEAT** and **Cyc+AUNet+iBEAT+AUNet**. Red circles point to areas where segmentations have been improved by the second Attention UNet.

More detailed comparisons between **Cyc+AUNet+iBEAT** and **Cyc+AUNet+iBEAT+AUNet** are visualized in **Fig. 4**. Adding a second Attention UNet at the end to train a segmentation model on the real 6-month T1w and T2w images improved the accuracy of WM segmentation in small structures, and this is highlighted in **Fig. 4** with the red circles.

## 4    Discussion

In this work, we utilized neonatal annotation information and a DA technique to develop the first 8-tissue segmentation pipeline for six-month-old infant brains. To satisfy the variability in modalities between individuals, we prepared the segmentation models for T1 modality only, T2 modality only and both T1, T2 modalities together. When leveraging both T1w and T2w information together, the model achieved the highest performance with a DICE score of 0.92, an HD95 of 1.6, and an ASSD of 0.42. When using a single modality as input, the performance decreased but not significantly (two-tailed t-test, $p=0.26$), obtaining a DICE of 0.89, an HD95 of 1.54 and an ASSD of 0.48 on the T1 modality, and a DICE of 0.90, an HD95 of 1.55 and an ASSD of 0.46 on the T2 modality.

When we incorporated the longitudinal registration algorithm into the pipeline, it resulted in less noisy and more accurate WM, GM, and CSF segmentations. However, it did not outperform iBEAT's cortical and WM performance, likely due to the registration accuracy limited by the low tissue contrast and image quality of the 6-month images.

The second Attention UNet at the end permits the training of a segmentation model capable of performing 8-tissue segmentation on 6-month-old infant brain images. It leverages tissue contrast from both T1w and T2w modalities, such as the contrast between cortex and unmyelinated WM from T2w images and myelinated WM from T1 images. This approach results in higher performance compared to single-modality segmentation.

The primary limitation of this work is the lack of data from other collection sites to evaluate the model's performance and generalization ability. In future research, we aim to acquire unseen datasets with varying acquisition parameters from different collection sites to validate the model's generalization ability. The segmentation results will effort to characterize potential associations between brain tissue features and atypical neurodevelopment, such as underlying neural differences related to autism.

**Acknowledgments.** This work was supported by King's-China Scholarship Council (K-CSC) (grant number 202008060109). The results leading to this publication have received funding from the Innovative Medicines Initiative 2 Joint Undertaking under grant agreement No 777394 for the project AIMS-2-TRIALS. This Joint Undertaking receives support from the European Union's Horizon 2020 research and innovation programme and EFPIA and AUTISM SPEAKS, Autistica, SFARI. The authors acknowledge support in part from the Wellcome Engineering and Physical Sciences Research Council (EPSRC) Centre for Medical Engineering at Kings College London (grant number WT 203148/Z/16/Z), and the NIHR Maudsley Biomedical Research Centre at South London and Maudsley NHS Foundation Trust and King's College London. The views expressed are those of the authors and not necessarily those of the funders, the NHS, the National Institute for Health Research, the Department of Health and Social Care, or the IHI-JU2. The funders had no role in the design and conduct of the study; collection, management, analysis, and interpretation of the data; preparation, review, or approval of the manuscript; and decision to submit the manuscript for publication.

**Disclosure of Interests.** The authors declare that they have no known competing financial interests or personal relationships that could have appeared to influence the work reported in this paper.

# References

1. Courchesne E, Pierce K, Schumann CM, et al (2007) Mapping early brain development in autism. Neuron 56:399–413. https://doi.org/10.1016/j.neuron.2007.10.016
2. Andrews DS, Lee JK, Harvey DJ, et al (2021) A Longitudinal Study of White Matter Development in Relation to Changes in Autism Severity Across Early Childhood. Biol Psychiatry 89:424–432. https://doi.org/10.1016/j.biopsych.2020.10.013
3. Wang L, Nie D, Li G, et al (2019) Benchmark on automatic six-month-old infant brain segmentation algorithms: The iSeg-2017 challenge. IEEE Trans Med Imaging 38:2219–2230. https://doi.org/10.1109/TMI.2019.2901712
4. Li G, Wang L, Yap PT, et al (2019) Computational neuroanatomy of baby brains: A review. Neuroimage 185:906–925. https://doi.org/10.1016/j.neuroimage.2018.03.042
5. Wang L, Wu Z, Chen L, et al (2023) iBEAT V2.0: a multisite-applicable, deep learning-based pipeline for infant cerebral cortical surface reconstruction. Springer US

6. Sun Y, Gao K, Wu Z, et al (2021) Multi-Site Infant Brain Segmentation Algorithms: The iSeg-2019 Challenge. IEEE Trans Med Imaging 40:1363–1376. https://doi.org/10.1109/TMI.2021.3055428
7. Jay F, Renou J-P, Voinnet O, Navarro L (2017) Unpaired Image-to-Image Translation using Cycle-Consistent Adversarial Networks Jun-Yan. Proceedings of the IEEE international conference on computer vision 183–202
8. Bui TD, Wang L, Lin W, Li G (2020) 6-MONTH INFANT BRAIN MRI SEGMENTATION GUIDED BY 24-MONTH DATA USING CYCLE-CONSISTENT ADVERSARIAL NETWORKS. 359–362
9. Liu J, Liu F, Sun K, et al (2023) Adult-Like Phase and Multi-scale Assistance for Isointense Infant Brain Tissue Segmentation. MICCAI, pp 56–66
10. Makropoulos A, Robinson EC, Schuh A, et al (2018) The developing human connectome project: A minimal processing pipeline for neonatal cortical surface reconstruction. Neuroimage 173:88–112. https://doi.org/10.1016/j.neuroimage.2018.01.054
11. Makropoulos A, Gousias IS, Ledig C, et al (2014) Automatic whole brain MRI segmentation of the developing neonatal brain. IEEE Trans Med Imaging 33:1818–1831. https://doi.org/10.1109/TMI.2014.2322280
12. Oktay O, Schlemper J, Folgoc L Le, et al (2018) Attention U-Net: Learning Where to Look for the Pancreas
13. Balakrishnan G, Zhao A, Sabuncu MR, et al (2018) An Unsupervised Learning Model for Deformable Medical Image Registration. Proceedings of the IEEE Computer Society Conference on Computer Vision and Pattern Recognition 9252–9260. https://doi.org/10.1109/CVPR.2018.00964
14. Cardoso MJ, Li W, Brown R, et al (2022) MONAI: An open-source framework for deep learning in healthcare
15. Grigorescu I, Uus A, Christiaens D, et al (2020) Diffusion tensor driven image registration: a deep learning approach. International Workshop on Biomedical Image Registration Cham: Springer International Publishing 131–140. https://doi.org/10.1007/978-3-030-50120-4_13